\documentclass[twocolumn,times,tighten,twocolappendix,resetfootnote]{aastex63}
\usepackage{graphicx,amssymb,amsmath,amsthm,amsfonts,epsfig}
\hypersetup{colorlinks=true, citecolor=blue, linkcolor=red, urlcolor=red}
\usepackage{float}
\usepackage{graphicx}
\usepackage{multirow}
\usepackage{amsmath,amssymb,amsfonts}
\usepackage{amsthm}
\usepackage{longtable}
\usepackage{hyperref}
\usepackage[flushleft]{threeparttable}

\hypersetup{colorlinks,citecolor=blue,linkcolor=blue,urlcolor=blue}

\newcommand{\Rm}{\mathcal{R}_{m}}

\newcommand{\BB}{\boldsymbol{B}}

\newcommand{\bnabla}{\boldsymbol{\nabla}}

\newcommand\blfootnote[1]{%
  \begingroup
  \renewcommand\thefootnote{}\footnote{#1}%
  \addtocounter{footnote}{-1}%
  \endgroup
}

\begin{document}

\title{Late-blooming magnetars: awakening as long period transients after a dormant cooling epoch}

\author{Arthur G. Suvorov$^{\dagger}$}
\affil{Departament de F\'isica, Universitat d’Alacant, Ap. Correus 99, E-03080 Alacant, Spain}
\affil{Theoretical Astrophysics, IAAT, University of T{\"u}bingen, T{\"u}bingen, D-72076, Germany}

\author{Clara Dehman$^{\dagger}$}
\email{clara.dehman@ua.es}
\affil{Departament de F\'isica, Universitat d’Alacant, Ap. Correus 99, E-03080 Alacant, Spain}

\author{Jos{\'e} A. Pons}
\affil{Departament de F\'isica, Universitat d’Alacant, Ap. Correus 99, E-03080 Alacant, Spain}

\begin{abstract}
\noindent Long-period transients are an elusive class of compact objects 
uncovered by radio surveys. While magnetars are a leading candidate 
for those sources that appear isolated, several observational properties challenge the established evolutionary framework: (i) low quiescent X-ray luminosities, (ii) $\sim$~hour-long rotational periods, and (iii) highly-variable radio flux. It is shown via magnetothermal modelling that, if electric currents thread the fluid core at the time of crust freezing, the neutron star remains multiband silent for an initial period of approximately 0.1~Myr while cooling passively. Once the crust becomes cold enough, the Hall effect begins to dominate the magnetic evolution, triggering crustal failures that inject magnetospheric twist that initiates radio pulsing while depleting rotational kinetic energy from an already-slow star. Depending on where electric currents circulate, such `late-blooming' magnetars manifesting as long-period transients may thus form a distinct branch from soft gamma repeaters and anomalous X-ray pulsars.
\end{abstract}


\section{Introduction}

\blfootnote{$\dagger$~These authors are to be considered equal contributors. }

Radio surveys are steadily revealing a Galactic population of astrophysical bodies termed long period transients (LPTs) \citep{hw23,caleb24,dong24}. 
Although a robust classification for this burgeoning class is yet to emerge, coherent radio pulsations from $\gtrsim$~15 LPTs have been detected since the discovery of GLEAM-X J162759.5--523504 \cite[henceforth GX J1627;][]{hw22}.
Optical spectroscopy has established that some LPTs are white dwarfs with type M companions \citep{rui24,hw24}. 
However, whether \emph{all} LPTs contain dwarfs remains debated, as neutron stars -- particularly magnetars -- are compelling candidates for those that lack binary signatures \citep{suvm23,lyman25} or exhibit interpulse \citep{lee25} phenomena.

The neutron-star hypothesis has garnered further support following the discovery  of DART/ASKAP J1832--0911 \cite[DA J1832;][]{li24,wang24}. 
This transient, with a period of $P \approx 44.2$~min, was detected simultaneously in the radio and X-ray bands by a host of instruments between December 2023 and September 2024, a discovery enabled by its serendipitous alignment with the supernova remnant G22.7--0.2.
This marked the first instance of such multiwavelength activity -- characteristic of transient magnetars that become radio-loud during outbursts \citep{tur15,cam16} -- in a LPT. 
The high radio luminosity (on-pulse $L_{\rm rad,max} \sim 10^{32} \text{ erg s}^{-1}$) and linear polarization ($\gtrsim 80\%$) of the source also disfavours a dwarf, as these features are difficult to explain with binary-instigated mechanisms \citep{mel17,qu25}.

Aside from the X-ray burst observed from DA J1832, \emph{isolated} LPTs share several features which have thus far eluded theoretical models.
They are: (1) cold --- there have been no persistent, soft X-ray detections to date;
(2) slow --- with periods spanning several minutes \citep{dong24} to a few hours \citep{lee25}; and 
(3) irregularly active --- often being radio-loud for only short windows within any given observational campaign \citep{hw22,caleb24}. 
Reconciling all of these properties within a single framework has proven challenging: magnetars are categorically hot, young, and with spin periods $P \lesssim 12$~s \citep{mpm15,coti18}.

Magnetar phenomena are attributed to their strong magnetic fields \citep{td95,td96}. 
They are observationally defined by explosive eruptions in the form of bursts or flares, triggered by the redistribution of magnetic flux as the field evolves. 
Joule dissipation partially counteracts neutrino cooling, and advanced models
incorporating this effect successfully reproduce the observed high luminosities \citep{vigano2013,coti18} and outburst rates \citep{gog00,gog01} as a function of age and magnetic intensity \cite[see][for a review]{pdv25}. 
The success of these magnetothermal models hinges on the proviso that electric currents circulate primarily within the crust of the star. 
This is because magnetic dissipation (and hence heating) is more effective there owing to the shorter length scales and presence of highly-resistive layers \citep{pons2013}. 
By contrast, simulations of neutron stars with fields of strength $\sim 10^{14}$~G and `core-threading' (CT) currents exhibit essentially passive cooling \citep{deh23b} and predict little activity within the first $\gtrsim$~kyr of life \citep{deh20}. 
Thus, observational considerations suggest that magnetars are born with magnetic configurations in which a substantial fraction of the sustaining electric currents reside in the solid crust, i.e. are predominantly crust-confined (CC).
The CC/CT distinction should not be understood as a strict dichotomy, but rather as one end of a continuum of possible current distributions, ranging from configurations confined to the outer layers of the crust to those extending progressively deeper into the stellar interior. 
The apparent prevalence of CC configurations may be explained in two broad ways. One possibility is that magnetars are born with magnetic field configurations in which most of the sustaining currents are already concentrated in the crust\footnote{A scenario explaining magnetar formation for a CC field configuration that accounts for the chiral magnetic effect is presented in \citet{dehman2025}.}. Alternatively, a substantial fraction of the magnetic flux may initially thread the core but be expelled to the crust at early times, rapidly establishing a crust-dominated configuration that can power magnetar activity on observable timescales \citep{brans25,lan24b}.

Magnetars typically slow down after outbursts \cite[e.g.,][]{arch17,borg21,low23}. 
This implies that some fraction of the energy that powers bursts and flares also goes towards decreasing the rotational kinetic energy. 
Because magnetic decay restricts high-energy activity to early evolutionary stages in CC models however, magnetar spin periods are limited to at most $\sim 20$~s during their active life \citep{pons2013}.

On the other hand, if outbursts persist at late times, albeit typically less frequent and less energetic in CT configurations \citep{pg07,vigano2013}, the associated spindown enhancement could still enable such stars to evolve into ``late-blooming magnetars'' with spin periods approaching $\sim$hour-long values \citep{ben20,ben23}.

In this work, we present a unified framework to explain the origin of isolated LPTs within the broader population of strongly magnetised neutron stars. We show that, depending on the regions in which electrical currents predominantly circulate, two distinct evolutionary pathways emerge: one leading to canonical magnetars (CC), and the other to isolated LPTs (CT).

Our approach combines three main ingredients. First, we perform long-term ($\sim$Myr) 2D magnetothermal simulations with realistic microphysical inputs, which self-consistently determine the thermal and magnetic evolution of the star, using the latest version of our 2D code \citep{vigano2021}. Second, the resulting magnetic configurations are used as inputs to a semi-analytic treatment of the rotational evolution, which is evolved separately in order to connect the magnetothermal histories to observable spin periods and spindown rates. Third, we employ phenomenological prescriptions for magnetically induced crustal failures, emission efficiencies, and activity thresholds, allowing us to link the evolutionary models to observable outbursts, radio activity, and X-ray phenomenology.

Together, this framework directly connects magnetothermal evolution to the characteristic periods, activity windows, and high-energy emission of LPTs, reproducing the three key observational properties of isolated LPTs. We begin in Sec.~\ref{sec:mtmain} by describing the magnetothermal evolution scheme and the role of current distribution in setting the evolutionary track. In Sec.~\ref{sec:outbursts}, we outline the phenomenological criteria used to estimate outburst rates and radio activity for each pathway, while Sec.~\ref{sec:sd} presents the associated rotational evolution and implications for enhanced spindown. We conclude in Sec.~\ref{sec:discussion} by discussing the implications of this model for future observations and for the existing population of LPTs.

\section{Magnetothermal evolution} \label{sec:mtmain}

\subsection{Crustal evolution} 

Within minutes after birth, nuclei in the outer layers of a neutron star freeze and form a solid crust \citep{yakovlev2004,potekhin2015}. Relativistic electrons within the crust carry a current that advects field lines through the Hall effect, generating small-scale structures that are especially susceptible to Ohmic dissipation \citep{mpm15,deh23_MT,dva23}. 
As hydrodynamical timescales are much shorter than either of these processes, the dynamics in the crust are well-described by the `electron-MHD' induction equation,
\begin{equation}
 \frac{\partial \BB }{\partial t} = -\bnabla \times 
    \left\{ \eta \left[ \bnabla \times \BB +  \Rm (\bnabla \times \BB) \times \hat{\BB} \right] \right\}.
   \label{eq: induction equation} 
\end{equation}
In the above, $\eta$ is the magnetic diffusivity and $\Rm$ is
the key plasma parameter -- analogue to the magnetic Reynolds number in resistive magnetohydrodynamics -- that dictates the evolutionary track. 
It is defined as the ratio between Ohmic ($\tau_{\Omega}$) and Hall ($\tau_{\rm H}$) timescales, viz.
\begin{equation} \label{eq:reynolds}
    \Rm = \frac{\tau_{\Omega}}{\tau_{\rm H}} = \frac{L^2/\eta}{4 \pi e n_e L^2/ c B} = \frac{c B}{4 \pi e n_{e} \eta},
\end{equation}
for magnetic length-scale $L$, speed of light $c$, elementary charge $e$, and electron number-density $n_{e}$. Note that the \emph{ratio} of timescales is independent of $L$.

Only when $\Rm \gtrsim 10^{2}$ do Hall-related phenomena -- where magnetic energy transfers between small and large scales -- operate \citep{pdv25,deh25}. 
This, combined with the fact that both the thermal and electrical diffusivities depend on the temperature and field strength, necessitates a coupled magnetothermal evolution. 
We solve the relevant heat-diffusion equation together with \eqref{eq: induction equation}, adopting the standard \citep{yakovlev2004,potekhin2015,deh23b} approach where the temperature at the crust-envelope boundary, $T_b$, is matched to a stationary envelope assuming radiative equilibrium. 
This links the thermal profile at the edge of the computational domain to the surface temperature, $T_s$, which determines the outgoing blackbody emission. 
The $T_s$-$T_b$ relation, which depends primarily on the average opacity and local gravity \citep{Gudmund83}, serves as an outer boundary condition for $T(t,\boldsymbol{x})$.
Technical details on the evolution scheme can be found in Appendix~\ref{sec:numerics}.

Several evolutionary pathways can lead to neutron stars with magnetic fields that are significantly stronger in the crust \citep{tf16,barr22,brans25} or even absent entirely \citep{lan25,lan24b} from the core. 
For such stars (i.e., those called CC in this paper), 
core-field evolution is irrelevant, and the above equations fully describe the system given a background stellar model.
Throughout this paper, we adopt the BSk24 equation of state \citep{pearson2018} for a `canonical' star with a mass of $M = 1.4 M_{\odot}$, radius $R = 12.426$~km, and crustal thickness of $0.86$~km.

\subsection{Core evolution} \label{sec:core}

\begin{figure*}
\centering
  \includegraphics[width=0.8\textwidth]{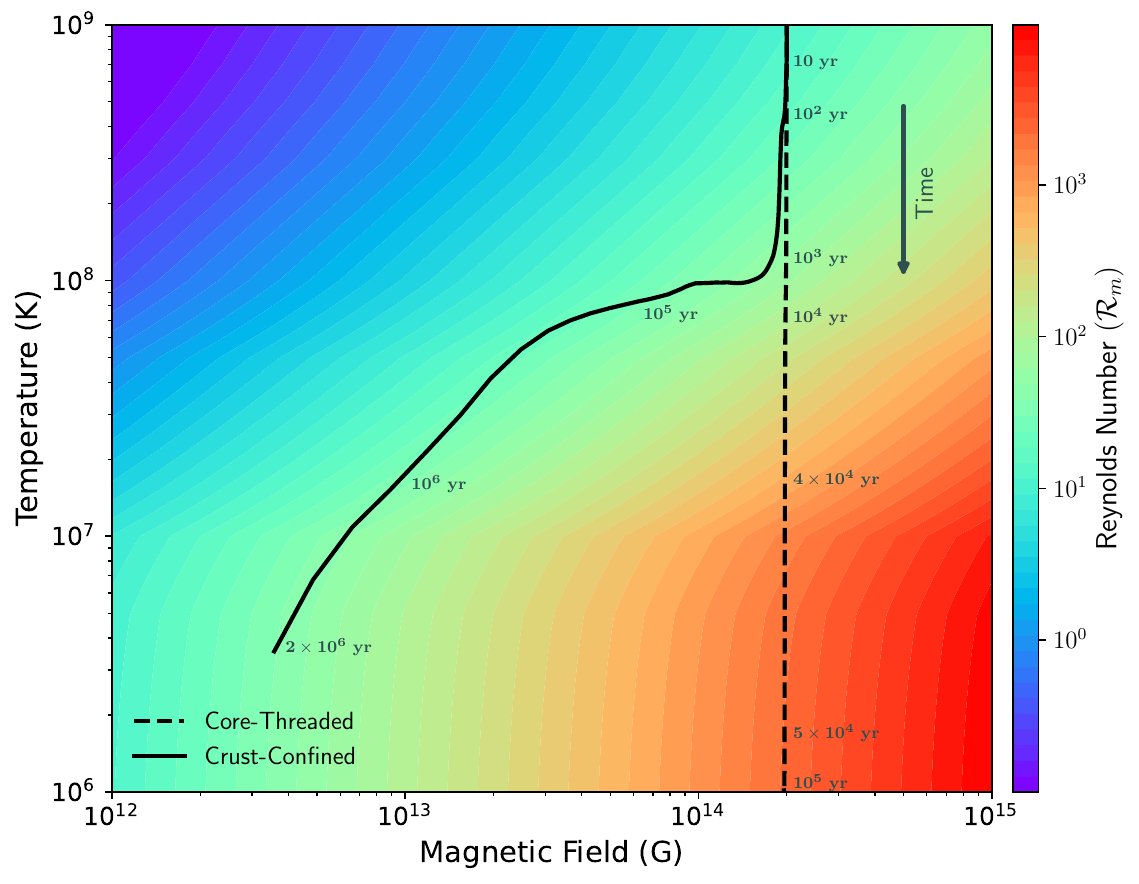}
  \caption{Evolutionary tracks, in the $T_b$-$B$ plane, for representative CC (solid) and CT (dashed) models. Stellar age is indicated by grey labels along the lines. The background colour contours show the magnetic Reynolds number ($\Rm$), calculated at a density of $ \rho \approx 5 \times 10^{10} $ g/cm$^3$ and polar latitude, with redder shades indicating greater $\Rm$. 
 }
  \label{fig:contours Rm}
\end{figure*}

If the core field is strong enough to be dynamically important, one must also specify how $\BB$ and $T$ evolve there. 
The core composition is unknown, but relative motions between charged particles and neutrals define an \emph{ambipolar velocity} which transports flux in a manner analogous \citep{gunsandroses92,vigano2021} to the Hall effect. 
Because these streaming motions are inherently hydrodynamical however, it is unclear which terms (if any) can be safely discarded in the multi-fluid equations describing momentum conservation. 
Stemming largely from the application of different approximation schema -- as modelling the full system over astrophysical timescales is impossible -- there are disagreements in the literature about the relevance of ambipolar diffusion \cite[see][for a review]{pdv25}. 
Even in `optimistic' scenarios \cite[e.g.][]{mor24} though, significant drifts likely take longer than a Myr for large-scale fields of strength $\sim 10^{14}$~G unless the core is very cold and the direct Urca process operates \citep{pass17}.

We neglect ambipolar drift and evolve the magnetic and thermal fields of stars with `moderate’ fields over Myr timescales. In CT models, the core field is formally evolved using the same induction equation as in the crust to ensure a smooth match at the crust–core interface. However, the effective Reynolds number in the core is very small and Hall dynamics are suppressed \citep{gunsandroses92}, so the core field remains essentially frozen over the timescales considered. Therefore, the core evolution has a negligible impact on the crustal dynamics and observable quantities, and could equivalently be replaced by a fixed-core boundary condition. Our approach is employed only for numerical convenience: it allows for a smooth matching of the magnetic field across the crust–core interface and avoids potential artifacts at the boundary. The currents are essentially chained to the core for CT models.

\subsection{Impact of the current distribution} \label{sec:jcirc}

We consider two representative examples throughout, featuring poloidally-dominated, large-scale magnetic fields with an average strength of $\gtrsim 10^{14}$~G that are practically identical with respect to physical parameters (total magnetic energy, mass and radius, multipolarity, and poloidal-toroidal partition). 
The key distinction lies in the initial location of electric currents (CC or CT), highlighting the role of circulatory motions.

Representative snapshots of the magnetic field evolution and associated crustal failures from the underlying 2D magnetothermal simulations, for both CC and CT configurations at the two fiducial field strengths considered here, are presented and discussed in detail in \cite{deh20}, to which we refer the reader for more information.

Fig.~\ref{fig:contours Rm} shows their respective, numerically-determined $T_{b}$-$B$ tracks. 
The overlaid colours depict the magnetic Reynolds number \eqref{eq:reynolds} at polar latitudes in the crust with both, initially-hot models beginning in the top-right of the diagram. 
The CT track (dashed line) descends vertically, reflecting cooling with minimal magnetic activity until $\lesssim 0.1$~Myr ages when $\Rm$ becomes large. 
By contrast, the CC model (solid) diverts within $\sim 10^{2}$~yr from this dive, evolving toward the bottom-left as the field decays in unison with cooling. 
The shorter length scales in the crust combined with the greater resistivity there are responsible for this bifurcation, and is why CC models better explain the X-ray profiles of Galactic magnetars. 
The Hall effect is particularly prominent for CC cases at ages $10^{1}  \lesssim t/\text{kyr} \lesssim 10^{2}$ where a horizontal plateau is visible, indicating an approximate balance between the heating and cooling rates at the expense of magnetic energy \cite[e.g.,][]{deh23b}.  

\begin{figure*}
\centering
  \includegraphics[width=0.94\textwidth]{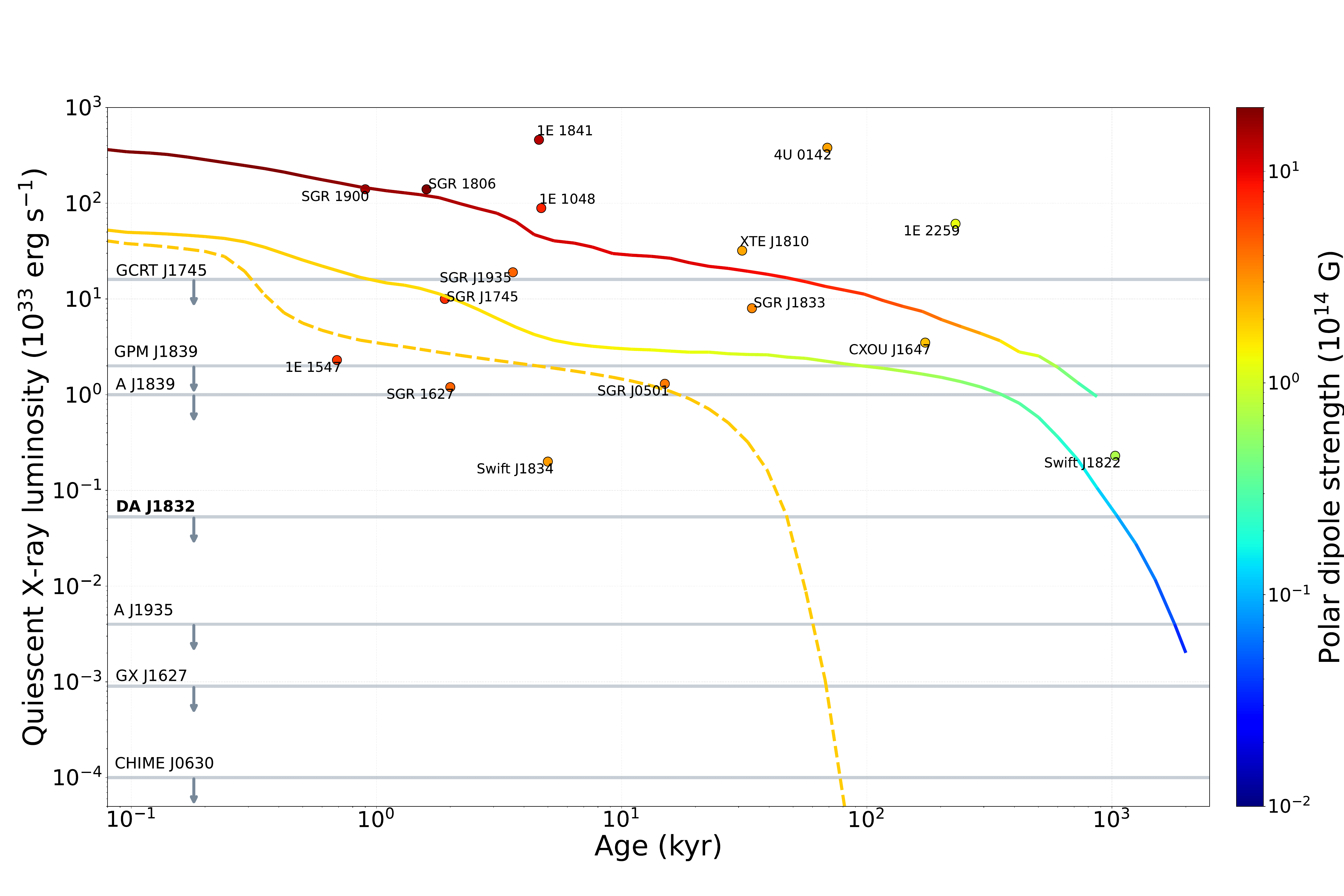}
  \caption{Comparison of simulated (lines) and observed (points) luminosities for magnetars and LPTs. Solid (CC) and dashed (CT) tracks represent the X-ray luminosity curves as a function of the `real' age as determined by the simulation. For the CC models, results are shown for two initial field strengths of $2 \times 10^{14}$~G (lower) and $2 \times 10^{15}$~G (upper). The color bar indicates the polar magnetic field intensity. Galactic magnetars are represented by circles as a function of their characteristic age. For seven isolated LPTs, upper limits on the X-ray luminosity are marked by the vertical arrows under horizontal lines as reliable age estimates are unavailable for these objects. 
  \label{fig: luminosity}
 }
\end{figure*}

Another way to visualise the impact of the current distribution is shown in Fig.~\ref{fig: luminosity}, depicting (redshifted) luminosity curves for CT (dashed) and CC (solid) cases overlaid with the observed populations of isolated magnetars \citep{coti18} and LPTs (see Appendix~\ref{sec:somelptproperties} for details on our sample). 
For the CT case, the quiescent X-ray luminosity drops below $L_{\rm X} \sim 10^{29} \text{ erg s}^{-1}$ within $\lesssim 0.1$~Myr. 
This reflects the vertical drop seen in Fig.~\ref{fig:contours Rm} with the star retaining its magnetic energy, providing a natural explanation for a cold and yet strong-field object (i.e., LPTs). 
For CC (solid) models, two cases with the same structure but different initial polar-dipole strengths ($2 \times 10^{14}$~G and $2 \times 10^{15}$~G) are shown, the larger of which better matches observations of Galactic magnetars. 
Because of Joule heating, the luminosity is too high to explain the observations of  GX J1627 or CHIME J0630+25 for either intensity even after $\gtrsim 2$~Myr.
These curves assume an envelope composed primarily of iron: 
a different composition or stellar mass could lead to a factor $\sim 10$ adjustment in $L_{\rm X}$ \cite[compare figures 2 and 4 in][]{deh23b} 
but, even at the lower end, CC models could not fit the coldness of LPTs before $\gtrsim$~Myr ages.
The field has decayed to $\ll 10^{14}$~G by that point though so as to disable outbursts and radio pulsing, as we explore next.

\section{Outbursts and late-time awakening} \label{sec:outbursts}

\begin{figure*}
\centering
  \includegraphics[width=\textwidth]{fig/histos_comb.png}
  \caption{Histograms of waiting times (left), energies (middle), and colatitudes (right) of failure events counted during evolution for the representative CT (top panel) and CC (bottom) models up to ages of 2~Myr and 1.375~Myr, respectively. Maximum-likelihood and method of moments fits, log-normal for both sets of waiting times, are overlaid with uncertainties. For the energy distributions, a gamma (log-normal) fit best represents the simulation output for the CT (CC) model.}
  \label{fig:ct4dists}
\end{figure*}

In this section, we investigate the implications of the magnetothermal evolution described above for magnetar outbursts and late-time activity. 
Our aim is not to model the detailed microphysics of crustal failure or emission processes from first principles, but rather to use physically motivated, semi-quantitative prescriptions to connect the evolution of simulated magnetic stresses to observable activity. 
Specifically, we identify crustal failure events within the 2D magnetothermal simulations, estimate their waiting times, energetics, and locations, and assess their consequences for spin-down, radio activation, and X-ray emission. This approach allows us to compare the qualitative and statistical differences between CC and CT evolutionary tracks, and to evaluate whether rare, low-level outbursts at late times can account for the observed phenomenology of LPTs.

The main driving idea is that, during a Hall-dominated epoch, flux waves of large amplitude can drift into malleable regions of the outer crust. 
The otherwise-rigid ion lattice is sheared as a result, and a mechanical strain is built up \citep{td95,thom02}. 
If a critical strain threshold is breached, a crustquake or some type of \emph{failure} can occur that liberates magnetoelastic energy \cite[e.g.,][]{lg19,deh20}. 
By tracking stresses throughout a given simulation we identify the failure sites.
In particular, consider the (traceless) Maxwell stress tensor,
\begin{equation}
   M_{ij}= \frac{1}{4\pi}\left(B_i B_j - \frac{1}{3} B^2 \delta_{ij}\right).
   \label{eq: traceless maxwell tensor}
\end{equation}
According to a von Mises criterion under a Hookean stress-strain relation, the crust fails when the absolute value of the deviation from equilibrium exceeds some threshold, 
\begin{equation} \label{eq:deltam}
    \Delta M = |M_{ij}-M_{ij}^{\rm eq}| \geq \sigma^{\rm max},
\end{equation}
for local stress limit $\sigma^{\rm max}$. 
These stresses are tracked throughout the simulations with event properties recorded when the inequality \eqref{eq:deltam} is met at any given spatial location and time step ($\sim 0.1$~yr). 
Note that, in many previous studies \citep{perna11,pons11,deh20}, the full stress tensor was used rather than the traceless component \eqref{eq: traceless maxwell tensor}. 
The latter is arguably more appropriate for predicting failures via deviatoric stresses \citep{lg19} from a continuum mechanics perspective; 
the adjusted criterion used here is more \emph{conservative} as magnetic pressures cannot directly contribute to overstraining (the trace of the full tensor is twice the magnetic pressure). 
A quantitative comparison between several different failure criteria is provided in Appendix~\ref{sec:failures}.

To determine the energetics of an event, we integrate the square of the relative difference between the magnetic field at the failure time and the reference state, 
defined as the field at the start of the current equilibrium epoch (i.e., defining $M_{ij}^{\rm eq}$ just after the previous failure). The critical stress threshold for crustal failure remains somewhat uncertain, as does the actual failure mechanism itself. 
If the crust behaves as a body-centred crystal though, molecular dynamics \citep{chugunov10} simulations indicate that
\begin{equation}
    \sigma^{{\rm max}} \approx \bigg( 0.0195 - \frac{1.27}{\Gamma-71} \bigg)n_i\frac{Z^2 e^2}{a},
    \label{eq: sigma max}
\end{equation}
for Coulomb parameter $\Gamma$ (electrostatic-to-thermal energy ratio), atomic charge $Z$, ion-number density $n_i$, and where $a$ is the typical inter-nuclear distance ($\propto n_i^{-1/3}$). 
The affected region readjusts to a new equilibrium, resolidifies, and then resumes its elastic response, accumulating stress until another rupture \citep{perna11,pons11}. 
Failures can avalanche as adjacent regions exceeding a threshold,
\begin{equation} \label{eq:threshold}
\Delta M \geq \epsilon \sigma^{\rm max}.
\end{equation} 
become unstable at sites determined by the crests of Hall or thermoplastic waves \citep{bell14}. In previous works \cite[e.g.,][]{deh20}, values in the range $0.8 \lesssim \epsilon \lesssim 1$ are typically used; we fix $\epsilon = 0.9$. 

Relevant findings for our two representative CT (top) and CC (bottom) cases are given in Fig.~\ref{fig:ct4dists}. 
We show the waiting time distribution of events (left), the energy released per failure (middle), and the colatitudes of failure sites (right). 
The CC waiting-time distribution follows a log-normal shape, consistent with prolific magnetars like SGR 1806–20 \citep{gog00} and SGR 1900+14 \citep{gog01} where ample burst statistics are available. 
This is expected as CC models are generally proposed to explain the properties of soft gamma repeaters and anomalous X-ray pulsars. 

\subsection{Radio activity cycle}

Although the exact nature of failures is poorly understood, because of the immense hydrostatic pressures it is thought that, rather than shattering like glass as brittle material would, failed regions of the crust deform plastically \citep{bell14,lg19}. 
Fluid circulating within `plastic islands' surrounding failure sites constitutes an effective longitudinal velocity which, via induction, produces an electric field that works to accelerate charges to high Lorentz factors. 
This initiates pair cascading and can trigger radio activity for relatively modest fields \cite[$\gtrsim 10^{14}$~G;][]{coop24}. 
One can think of the plastic flow as a localised circulation which dominates over rigid rotation for LPTs. 
In particular, radio-activation theory demands a minimum voltage drop ($\Delta V \propto B P^{-2}$) which is difficult to achieve for LPTs unless  $B \gg 10^{14}$~G because $P$ is so large \citep{cr92,suvm23}. 
Therefore, the star can only become radio-loud when a failure occurs and a plastic flow is initiated \cite[see][]{coop24}.
The relative `activity window' set by the duration of a plastic flow and frequency of failures defines a physical \emph{duty cycle}, $D^{\rm p}$, which can be quantified. 

For CT cases -- the proposed LPT track -- the waiting time distribution peaks at $\sim 1$~yr, and thus if motions circulate in a plastic island for $\sim 6$~months (as anticipated from the radio continuity of DA J1832 and other LPTs), we estimate an upper limit of $D^{\rm p}_{\rm max} \sim 0.5$. 
On the other hand, given the narrowness of plastic islands, an observer will catch the beam only after events that are favourably positioned with respect to their line of sight. 
Even if the \emph{physical} duty cycle is large therefore -- in the sense that a cone of radiation is emitted perpendicular to the plastic island that forms after each failure --  the \emph{observed} duty cycle, $D$, is reduced by a solid angle \citep{coop24}. 
For example, if the star is positioned such that only failures occurring at a narrow range of co-latitudes surrounding $\theta \sim 0.3$~rad are visible to a given observer, the right panel of Fig.~\ref{fig:ct4dists} indicates that only 1 in $\gtrsim 10^{2}$ events may reveal in radio (i.e., $D \sim 0.5\%$).  
Such a configuration would explain the observed \citep{dong24} value of $D \sim 0.6\%$ from CHIME J0630+25 and other isolated LPTs which display duty cycles of order $\sim 1\%$ (see Table~\ref{tab:isolated LPTs} and Appendix~\ref{sec:somelptproperties}). 
While it is difficult to falsify the model based on $D$ values because LPT viewing geometries are unconstrained, one prediction of the model is clear: X- and radio-band activity should go together since both originate from the same event.

\subsection{X-rays and interpulses}

Our CT energy distributions imply dim outbursts: if $\sim 1\%$ of the failure energy converts to observable X-rays \citep{blaes89,brans20}, most bursts would fall below detection thresholds. 
This may explain why only DA J1832 has shown coincident X-ray and radio activity, reflecting observational limitations rather than an intrinsic rarity of outbursts.
Moreover, many LPTs were identified only in archival data, and thus X-ray follow-up some years after radio activation is inappropriate to assess the scenario proposed here. 
In contrast, high-cadence monitoring of the relatively active \citep{hw23,men25} LPT GPM J1839--10 ($D \sim 25\%$) would be especially useful to test the `late-bloomer' picture.

The colatitude distribution of CT failure sites is bimodal at polar angles. This is because the initial field is approximately north-south symmetric, 
and so Hall drift frequently induces failures at \emph{both} polar caps in rapid succession. 
This naturally explains interpulse observations \citep{lee25} of ASKAP J183950.5--075635 (ASKAP J1839): two diametrically opposed plastic islands form at roughly the same time and the two beams sweep past the observer with a phase offset of $\approx 0.5$.
The model therefore requires that LPTs are born with fields of low multipolarity.
If the field were more isotropic, there would be no preferred latitudes for failure sites (similar to the CC model) and interpulses would be improbable.
Polar latitudes in particular also explain why a significant phase overlap between the 1--10~keV flux and radio pulsing was observed from DA J1832, as we expect the pulsed component of the X-rays to be spatially coincident with the radio-emission zone if a given failure is responsible for both phenomena. 
Similar arguments were made by \cite{belo09} for events from XTE J1810--197, whose pulse profile resembles that of GPM J1839--10 \citep{men25}.

\section{Enhanced spindown} \label{sec:sd}
 
\begin{figure*}
\centering
  \includegraphics[width=0.492\textwidth]{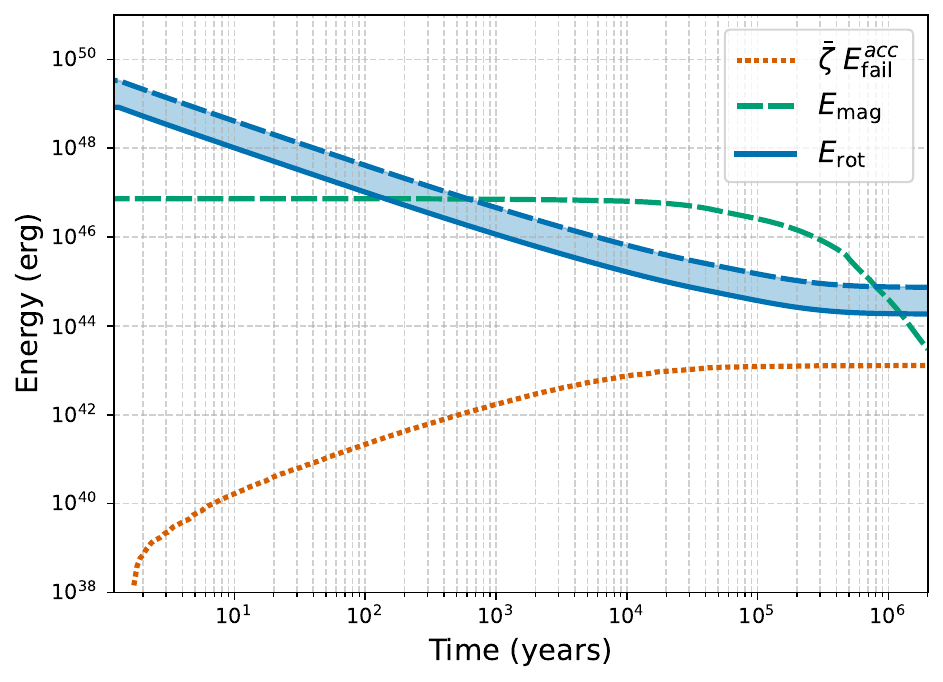}
  \includegraphics[width=0.492\textwidth]{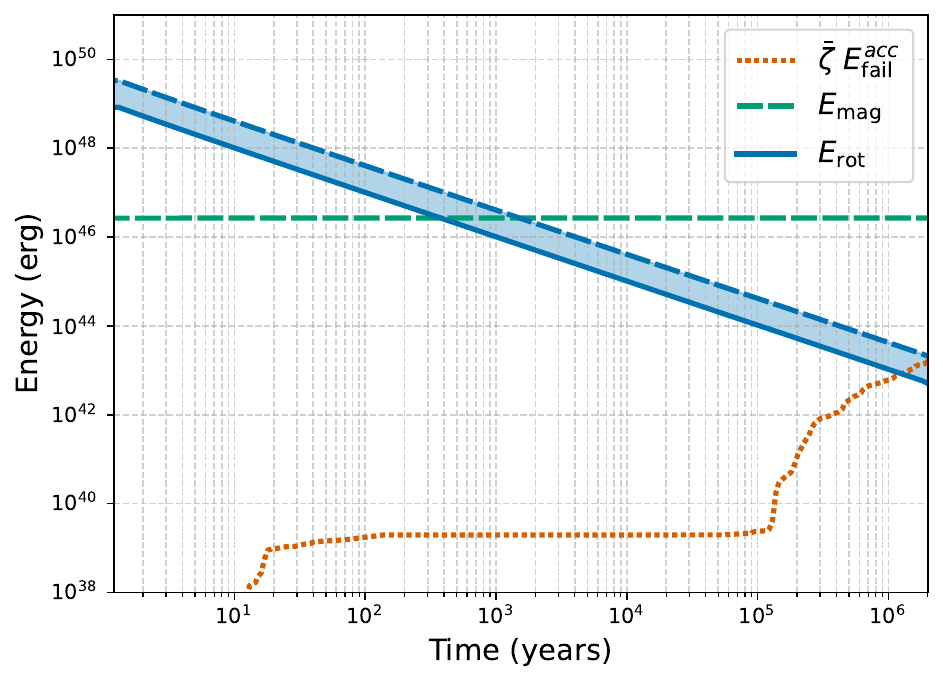}
  \caption{Time evolutions of the normalised ($\bar{\zeta} = 10^{-3}$) cumulative failure (dotted red), magnetic ($E_{\mathrm{mag}}$, dashed green), and rotational ($E_{\mathrm{rot}}$, blue) energies for representative CC (left) and CT (right) cases. The solid (dashed) blue curve corresponds to the spindown law \eqref{eq:braking} for an orthogonal (aligned) rotator.
  }
  \label{fig: model 4 CT CC}
\end{figure*}
 
Neutron stars gradually spin down due to electromagnetic braking torques.
For a star with a (time-dependent) polar-dipole strength $B_{p}$, inclined relative to the rotation axis by an angle $\alpha$, the period evolves according to\footnote{Note that we ignore general-relativistic corrections, which \emph{accelerate} spindown for a given $B_{p}$, to provide a conservative scenario \cite[see, e.g.,][]{ssp25}.} \citep{spit06}
\begin{equation} \label{eq:braking}
P \dot{P} \approx \frac {2 \pi^2 B_{p}^2 R^{6}} {3 c^3 I_{0}} \left(1 + \sin^2 \alpha \right),
\end{equation}
for moment of inertia $I_{0}$. However, if the external field anchored to the crust is twisted by failures, the resulting increase in magnetic pressure will open up more lines that carry away angular momentum \citep{thom02,belo09}. 
This will occur in stages, with a more `glitch-like' jump in $P$ caused by the abrupt release followed by a gradual deceleration as the twist dissipates. 
Evidence for a timing glitch in the LPT CHIME J0630 supports this picture \citep{dong24}; sadly, the source was not monitored in X-rays at the time to compare with the model.

A relationship between failure energies and rotational kinetic-energy losses can be quantified phenomenologically by an \emph{efficiency}, $\bar{\zeta}$, viz.
\begin{equation} \label{eq:effic}
   \Delta E_{\rm rot} = -\bar{\zeta} E_{\rm failure},
\end{equation}
where $E_{\rm rot} = 2 \pi^2 I_{0}/P^2$. 
An estimate for the efficiency can be made by first comparing theoretical expectations for how much of the failure energy escapes as X-rays, thought to be of order $\sim 1\%$ \citep{blaes89,brans20}. 
Observations of spindown losses \citep{borg21,low23} beyond the baseline model \eqref{eq:braking} following outbursts from radio-loud magnetars with pulse profiles most similar to that of LPTs indicate that $\Delta E_{\rm rot} \sim -0.1 E_{\rm burst}$. 
The overall efficiency is the product of these two fractions, and so we consider $\bar{\zeta} = 10^{-3}$ as representative (see Appendix~\ref{sec:sdcontext} for further details on this choice).

Fig.~\ref{fig: model 4 CT CC} presents time evolutions for the cumulative failure energy (dotted; scaled by $\bar{\zeta} = 10^{-3}$), magnetic energy (dashed), and rotational kinetic energy (thick band) for our representative CC (left panel) and CT (right) models.  
Note that $B_{p}$ is almost constant over the first $\sim 2$~Myr for the CT configuration (green curve in Fig.~\ref{fig: model 4 CT CC}), so that the combination $P \dot{P}$ appearing within equation \eqref{eq:braking} is also roughly constant.
For CC cases, the polar field-strength decays and the baseline $E_{\rm rot}$ effectively plateaus after a $\sim$~Myr. 
This decay also arrests the stress build-up process after $\sim 10$~kyr and so failures, which were otherwise regularly excited, halt. 
Towards the end of the CC simulation we have $\bar{\zeta} E^{\rm acc}_{\rm fail}/E_{\rm rot} \sim 10^{-2}$, 
implying that adolescent activity cannot allow the system to mature with a long period since $E_{\rm rot}$ flattens and failures no longer occur due to field decay.

While CC models dominated by small-scales and with stronger fields can better explain X-ray observations of Galactic magnetars (Fig.~\ref{fig: luminosity}), they fare worse in terms of explaining LPTs. 
This is because, even though $\Rm \propto B$ is itself independent of the system length-scale, both the Hall and Ohmic times scale as $L^2$ which is smaller for multipoles (equation \ref{eq:reynolds}). 
This prevents \emph{any} CC model from reaching LPT-like periods even if $P \sim 10$~s earlier in life due to the stronger field \citep{pons2013}. 
In CT simulations, failures only start taking place en masse when $P \sim 10$~s and $\Rm \gtrsim 100$. 
Around $\gtrsim 1$~Myr, the cumulative failure energy becomes comparable to the rotational energy for $\bar{\zeta} \approx 10^{-3}$, 
indicating that injected twists could effectively drain the entire rotational kinetic energy reservoir. 

Up to this point, we have discussed rotational evolution primarily in terms of energetic considerations. We now introduce a more explicit, physically motivated prescription for spindown enhancement associated with crustal failure events. In this model, each failure temporarily modifies the large-scale magnetic configuration by increasing the effective dipole moment entering the braking law in Eq.~\eqref{eq:braking}. Specifically, when a failure is triggered, the dipole moment is assumed to increase over a short rise time by a factor proportional to the energy released in the event.

The amplitude of this enhancement is parameterised 
by the dimensionless efficiency factor, $\zeta$, 
defined in Eq.~\eqref{eq:effic}. The normalisation of this factor is constrained by observations: we require that the resulting spindown rate satisfies $\dot{P} < 10^{-9}\mathrm{\,ss^{-1}}$, consistent with the observational upper limit inferred for DA~J1832 \citep{wang24}. Following the rise phase, the enhanced dipole moment is assumed to decay linearly back to its pre-event value, in rough agreement with the observed relaxation of magnetospheric twists following magnetar outbursts \citep{tur15}. The decay timescale is not chosen arbitrarily, but is fixed self-consistently by enforcing Eq.~\eqref{eq:effic} as an integral constraint on the total rotational energy extracted during each event (i.e., that the time-integrated energies match).

\begin{figure}
\centering
  \includegraphics[width=0.49\textwidth]{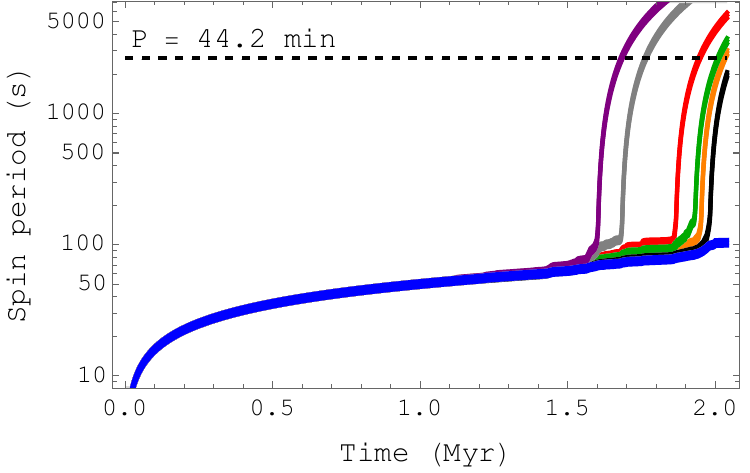}
  \caption{Spindown trajectories for the CT model with various efficiencies spanning the range $7 \leq \bar{\zeta}/(10^{-4}) \leq 14$ in increments of $10^{-4}$ (with $\bar{\zeta}$ increasing from left to right). The spin period of DA J1832 ($P \approx 44.2$~min) is overlaid by the dashed, horizontal line for reference. Models with $\bar{\zeta} \gtrsim 8 \times 10^{-4}$ are sufficiently slow by $t \sim 2$~Myr to match the observed pulse period of DA J1832.
  }
  \label{fig:ct_sd}
\end{figure}

Using this prescription, we generate ensembles of spindown histories by sampling the event energies and waiting times from the distributions shown in Fig.~\ref{fig:ct4dists}. The resulting rotational evolution for representative realisations is shown in Fig.~\ref{fig:ct_sd}, for mean efficiencies in the range $0.07\% \leq \bar{\zeta} \leq 0.14\%$, in steps of $0.01\%$. For $\bar{\zeta} \gtrsim 0.08\%$, the cumulative spindown enhancement induced by repeated failure events is sufficient to slow the star to spin periods of order $\sim$hour at Myr ages.
Finally, we note that modest variations in magnetic field strength between sources naturally modify both the event rate and the characteristic value of $\bar{\zeta}$. As discussed in detail in Appendix~\ref{sec:sdcontext}, this allows a wide range of late-time spin periods to be produced within the same framework, even among objects of comparable age.

\section{Discussion and future prospects} \label{sec:discussion}

We argue in this paper that the defining characteristics of isolated LPTs can be explained with a magnetar model provided that electrical currents circulate through the core (a l{\'a} CT). 
This condition alters the star's trajectory in the $T$-$B$ phase space, as illustrated in Fig.~\ref{fig:contours Rm} depicting the magnetic Reynolds number. 
Our simulations show that $\Rm \gtrsim 10^{2}$ at $\gtrsim 0.1$~Myr ages, whereupon the star is cold ($T_{s} \lesssim 10^{6}$~K) and too dim to be observable in X-rays (Fig.~\ref{fig: luminosity}).
At this age, it enters into a delayed Hall-dominated epoch such that it effectively reawakens \citep{pg07,vigano2013} as what we have called a `late-blooming magnetar' whereupon crustal failures (Fig.~\ref{fig:ct4dists}) start seeding both radio and X-ray activity. 
Critically, rotational energy is already low ($\sim 10^{44}$~erg; Fig.~\ref{fig: model 4 CT CC}) at this stage meaning that, if even a fraction ($\bar{\zeta} \sim 10^{-3}$; Fig.~\ref{fig:ct_sd}) of the failure energy goes into spindown, the star could reach hour-long periods. 
The radio duty cycle is tied to the failure rate in this picture; a prediction of the model is thus that of a positive correlation between periods and duty cycles in LPTs. 
This is supported by the (small sample-size) data as the lowest-$P$ source (CHIME J0630) also has the lowest duty cycle ($D \sim 0.6\%$). 
GPM J1839--10 is an anomaly however, and we require that its pole is face-on so that the \emph{observed} duty cycle is about half of the physical one ($D \approx D^{\rm p}/2$) or that it is in a binary \cite[see][for a discussion]{sdp25b}.

We find that failures cluster around the poles if the birth field is well-ordered, naturally explaining the latitude of the hotspot of DA J1832 \citep{wang24} and interpulses from ASKAP J1839 \citep{lee25}. We note, however, that this polar clustering is also influenced by the axisymmetric nature of the simulations, which favours the development of coherent large-scale structures, as reported in earlier 2D studies \citep{perna11}. 
Fully 3D models may broaden the latitude distribution of failure sites, although we expect polar regions to remain favoured when the large-scale field geometry is predominantly dipolar.

Low multipolarity also favours a fossil origin over a turbulent dynamo for the birth field \citep{barr22}. 
For example, recent 3D simulations  of core-collapse by \cite{varma23} for a static, magnetised $17 M_{\odot}$-star found that $0.4$~s after bounce the dipole field ($\approx 3 \times 10^{14}$~G) could make up $\approx 40\%$ of the RMS intensity ($\approx 7 \times 10^{14}$~G) at densities surrounding $10^{11} \text{ g cm}^{-3}$.
Promising progenitors are stars like $\xi^{1}$ Canis Majoris, a highly-magnetised $\sim 14 M_{\odot}$ variable revolving once every $> 30$~years \citep{schultz17}. 
If the star continues to slow at its current rate it would be static in $<$4~Myr, whereupon collapse a slow magnetar may form by magnetic-flux and angular-momentum conservation. 
This scenario fits best with the RCW 103 supernova remnant 1E 161348--5055, which exhibits a variable X-ray luminosity spanning $\sim 10^{33}$ to $\sim 10^{35}$~erg/s indicative of ongoing heating despite showing a period of many hours \citep{del06}. 
This object is difficult to explain with a CT model, but failures in a CC model could drain the entire reservoir after a few kyr \emph{if} it were born with $E_{\rm rot} \lesssim 10^{45}$~erg (Fig.~\ref{fig:ct4dists}). 

Mass-loaded winds could also lead to exponential increases in the spin period in a manner strongly resembling our Fig.~\ref{fig:ct_sd} \cite[see][]{hard99,ben20,ben23}.
While previous models of flare-induced spindown are phenomenologically similar to that proposed here, we emphasise that our predictions
(i) arise self-consistently from magnetothermal outputs, 
(ii) do not rely on global events, and 
(iii) only work in CT configurations for typical birth periods.
The second point is worth bearing in mind since, although extensive torque variability was observed in the lead-up and aftermath to the event, minimal changes were seen in $\dot{P}$ surrounding the giant flare from SGR 1806--20  \citep{youn17}. 
One way in which these scenarios could be confronted in future is through metal abundances: mass-loaded winds should invite $r$-process nucleosynthesis and hence enrich LPT nurseries \citep{thom25}. 

Another possibility proposed by \cite{lan24b} for producing isolated LPTs is that of a delayed Meissner expulsion.
This could resurrect a slow, CC magnetar by pushing flux into the crust. This scenario would, in essence, be represented by the left panel of Fig.~\ref{fig: model 4 CT CC} but with a lower `initial' $E_{\rm rot}$, so that when crustal failures begin to sap energy the star could reach $\bar{\zeta} E^{\rm acc}_{\rm fail} > E_{\rm rot}$. 
However, forcing huge fluxes into the crust should generate heat: this fits with 1E 1613 but is difficult to reconcile with the coldest LPTs. 
One may also anticipate a skew towards larger burst energies in this case (top panel of Fig.~\ref{fig:ct4dists}), which is tension with the low energy of the DA J1832 event and lack of visible outbursts from other LPTs. 
If a bright flare from a LPT is detected in future, such a scenario may be favoured since the maximum failure energy of CT models is of order $\sim 10^{41}$~erg (bottom panel of Fig.~\ref{fig:ct4dists}). 
Hall waves of high amplitude may also be launched into the crust as flux is thrust there \citep{brans25}, through a Meissner expulsion or otherwise. 
Notably, waves of the low-frequency, whistler variety may enduringly propagate in latitudinal directions such that the ambient poloidal field oscillates violently, instigating intermittent heating, anomalous torques, avalanche phenomena, and multiband transients. 
Ab initio modelling in this direction could pave the way for population synthesis studies of canonical vs. `late-blooming' vs. `reborn' magnetars.  

\section*{Acknowledgements}
We thank Adam Dong, Zorawar Wadiasingh, Yuri Levin, and Andrei Beloborodov for valuable discussions, and Alex Cooper for insights on pulsar activity driven by plastic flow. 
AGS acknowledges funding from the European Union's Horizon MSCA-2022 research and innovation programme ``EinsteinWaves'' under grant agreement No. 101131233, the Deutsche Forschungsgemeinschaft individual research grant 570901071, and the High Performance and Cloud Computing Group at the Zentrum f{\"u}r Datenverarbeitung of the University of T{\"u}bingen (Project ``AnqaGW'').
CD is supported by a Juan de la Cierva fellowship JDC2023‐052227‐I, funded by MINISTERIO DE CIENCIA, INNOVACIÓN Y UNIVERSIDADES, AGENCIA ESTATAL DE INVESTIGACIÓN, and UNIÓN EUROPEA (FSE+). 
We gratefully acknowledge the support provided by the Conselleria d'Educaci{\'o}, Cultura, Universitats i Ocupaci{\'o} de la Generalitat Valenciana through Prometeo Project CIPROM/2022/13, and from the AEI grant PID2021-127495NB-I00 funded by MCIN/AEI/10.13039/501100011033.

\section*{Author Contributions}
The first two authors of this work (C.~Dehman and A.G.~Suvorov) contributed significantly to the conceptualisation, implementation, model design, writing, and editing: to that degree it is with regret that lead authorship cannot be more perceptibly shared. Their contributions were distinct but approximately equal (e.g., A.G.~Suvorov handled spindown elements and C.~Dehman conducted the magnetothermal simulations and formulated the corresponding results presented in this manuscript, with ongoing mutual feedback). J. Pons was also directly involved in matters of editing, presentation, and physics discussions regarding the key properties of the model(s) and simulation setup(s).

\bibliography{ULPbib_N}

\appendix

\section{Numerical details on magnetothermal scheme} \label{sec:numerics}

Axisymmetric evolutions of magnetic fields are considered in this work, utilising a recent version of a 2D magnetothermal code \citep{vigano2021}. This code numerically integrates the induction \eqref{eq: induction equation} and heat-diffusion equations,
\begin{equation}
  c_{\rm V} \frac{\partial \left(\mathrm{e}^{\Phi} T \right)}{\partial t}
      - \bnabla\cdot[\mathrm{e}^{\Phi}
 \hat{\boldsymbol{\kappa}}\cdot\bnabla(\mathrm{e}^\Phi T) ] = 
           \mathrm{e}^{2\Phi} (Q_J - Q_\nu),
\label{eq:heatdiff}
\end{equation}
simultaneously within the neutron star interior. Here, $Q_J$ and $Q_{\nu}$ represent the heating power and neutrino emissivity, respectively,
with $c_{\rm V}$ being the specific heat, $\hat{\boldsymbol{\kappa}}$ being the thermal conductivity tensor, and $\mathrm{e}^{\Phi}$ being a metric factor \citep{vigano2021}. Note that in equations \eqref{eq:heatdiff} and \eqref{eq: induction equation}, general-relativistic effects are implicitly incorporated via the gradient operator. At the outer crust, defined numerically by a mass-density of $\rho = 10^{10} \text{ g cm}^{-3}$, we impose a current-free condition. While this is not strictly consistent with our spindown model, where magnetospheric twists should be injected following failure events, it should not alter the bulk statistics in a way that qualitatively changes our conclusions.

We assume an initial, isothermal temperature of $10^{10}$~K, typical of the end of the proto-stage when the star becomes transparent to neutrinos \citep{pons99}. 
Importantly, the outer layers (envelope and atmosphere) react on short timescales and quickly reach radiative equilibrium due to their low density. 
Simulating cooling across all layers up to the stellar surface would be computationally demanding, as it requires resolving vastly different timescales on the same numerical grid. 
Instead, we adopt a precomputed $T_{s}$-$T_{b}$ condition as detailed in the main text. We consider an iron envelope model, expected for isolated stars without a history of accretion. 
Note that an envelope composed of lighter elements will be \emph{hotter} at early times (i.e., before photon cooling begins to dominate over neutrino emissions) than one made primarily of iron \citep{deh23b}. 
For an object like 1E 1613 (or Galactic magnetars), an envelope composed of lighter elements may better fit the X-ray luminosity; see \cite{deh23b} for a detailed discussion. 

We incorporate superfluid and superconducting gaps for neutrons and protons, respectively, as they play a crucial role in cooling timescales by influencing the heat capacity and neutrino emissivity. 
While our model includes these effects, we do not alter the equations of motion to incorporate additional effects. For a comprehensive formulation of dissipative magnetohydrodynamic equations for finite-temperature superfluid and superconducting charged relativistic mixtures that account for particle diffusion, we refer to \cite{dom21}.
Thermal and electrical conductivities are computed using realistic data from the IOFFE repository\footnote{\url{http://www.ioffe.ru/astro/conduct/}}. For a complete discussion of all these microphysical ingredients, beyond that which we can adequately explore here, we refer the reader to \cite{potekhin2015}.

To describe the hydrostatics of the background state, we solve the Tolman-Oppenheimer-Volkov equations with the BSk24 equation of state (EOS), not accounting for (small) thermal corrections \citep{pearson2018}. This EOS includes consistent crust and core regions and matches well with observational constraints from gravitational waves from binary-mergers, maximum-mass observations of pulsars, fast-cooling constraints, phase-coherent X-ray spectra mapped by the Neutron Star Interior Composition Explorer Mission (NICER), and other channels \citep{marino24,suv24}.
At the magnetic field strengths and temperatures considered here, such a cold and unmagnetised EOS is an excellent approximation. A spherically-symmetric star is assumed, ignoring rotational oblateness and/or magnetic deformations to the mass-density. The above factors could, in principle, influence the evolution (at least at early times) and introduce complicated anisotropies into the crustal conductivity profiles, but such effects are expected to be small. 

\subsection{Crustal failures} \label{sec:failures}

\begin{figure*}
\includegraphics[width=\textwidth]{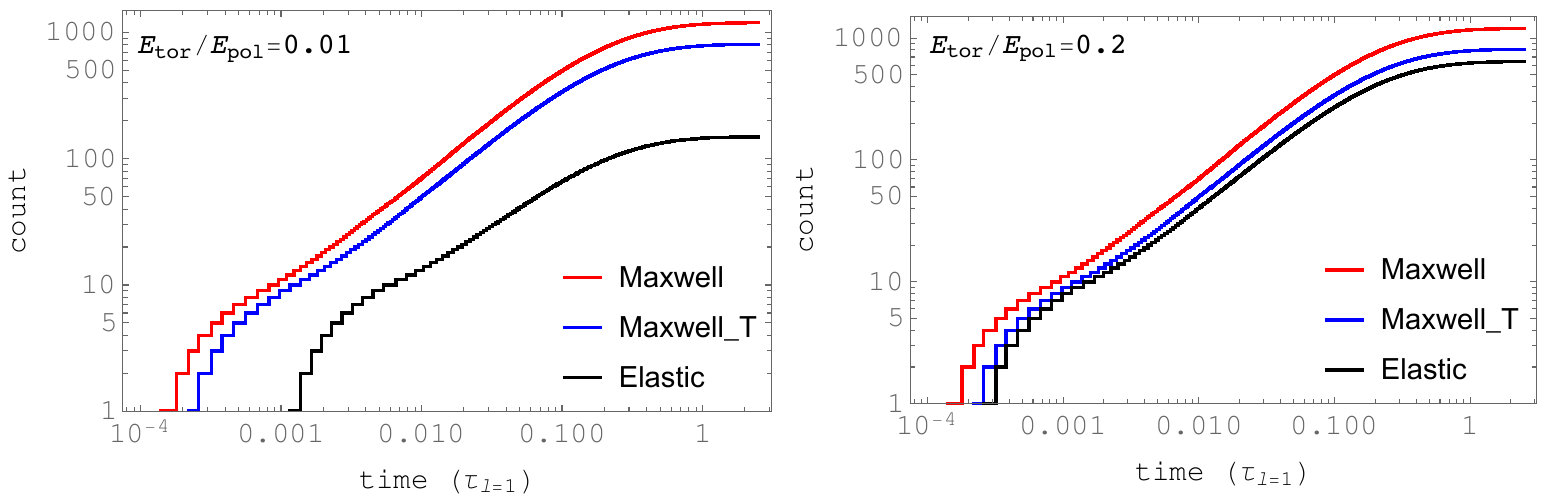}
  \caption{Comparison between event rates for a simple implementation of the failure problem using ``volume-averaged'' induction \eqref{eq:bfield} and a constant shear modulus as a function of time normalised to Ohmic time, with $\tau_{\rm H} = \tau_{\Omega}/2$. Within the figure legends, ``T'' stands for traceless.}
  \label{fig:comp}
\end{figure*} 

In the main text, a value of $\epsilon = 0.9$ appearing within equation \eqref{eq:deltam} is fixed.
In general, higher values increase event frequency but reduce affected areas, while lower values lead to larger but more infrequent failures. 
We do not expect that such changes would qualitatively adjust our findings here, since we are primarily interested in the global energetics of events, at least as far as spindown is concerned (see Fig.~\ref{fig: model 4 CT CC}). 
For deformed crystals, $\sigma_{\rm max}$ is expected to decrease and thus its use here implies an underestimate of the failure rate and overestimate of individual energies. 
While such adjustments are degenerate with $\epsilon$ to some degree, meaning that a lower threshold should roughly correspond to a more malleable crystal, it is difficult to quantitatively estimate how the two relate.
Note also that the threshold may instead \emph{increase} in `pasta' regions (i.e., nonspherical configurations of nuclei, as related to the inter-nuclear distance), although this is unimportant as the majority of failures found in our simulations occur in the outermost layers ($r \gtrsim 0.98 R$). 

Perhaps more importantly, \cite{brans25} note that by tracking Maxwell stresses (as in Sec.~\ref{sec:outbursts}), rather than mechanical ones directly, overestimates for event rates may be obtained. 
While modifying the code to compute Lagrangian displacements to more realistically estimate stresses lies beyond our current capabilities, we provide here a worked example that illustrates for mixed poloidal-toroidal fields of comparable strength the rate estimates are relatively insensitive to the exact criterion used.

As discussed by \cite{brans18} and others, the strong radial stratification in the crust permits only horizontal ($\xi_{r} = 0$) and incompressible ($\nabla \cdot \boldsymbol{\xi} = 0$) deformations because the hydrostatic pressure is much larger than the Coulomb energy density of the crystal. In the axisymmetric limit, the Lagrangian deformation thus reads
\begin{equation} \label{eq:axixi}
\boldsymbol{\xi} = \xi_{\phi}(t,r,\theta) \hat{\boldsymbol{e}}_{\phi},
\end{equation}
while incompressibility ensures that the elastic strain tensor can be written in the Hookean limit as
\begin{equation} \label{eq:elastic}
\boldsymbol{\tau} = \mu \left[ \nabla \boldsymbol{\xi} + (\nabla \boldsymbol{\xi})^{T} \right] /2,
\end{equation}
for shear modulus $\mu$. In the crust, the displacement $\boldsymbol{\xi}$ is determined by solving is the azimuthal component of the momentum-balance equations,
\begin{equation} \label{eq:balance}
   \nabla \cdot \boldsymbol{\tau} =  \left( \nabla \times \boldsymbol{B} \right) \times \boldsymbol{B} / 4 \pi.
\end{equation}
Following \cite{pg07} we write
\begin{equation} \label{eq:bpol}
\boldsymbol{B}_{\rm pol} = \nabla \times \left( - \boldsymbol{r} \times \nabla \Phi \right), \quad \boldsymbol{B}_{\rm tor} = - \boldsymbol{r} \times \nabla \Psi,
\end{equation}
for two generic functions $\Phi$ and $\Psi$ and position vector $\boldsymbol{r}$. Using an (axisymmetric) spectral decomposition of the form
\begin{equation}
\Phi = \sum_{\ell} \frac{\Phi_{\ell}(t,r)}{r} P_{\ell}(\cos \theta), \Psi = \sum_{\ell} \frac{\Psi_{\ell}(t,r)}{r} P_{\ell}(\cos \theta),
\end{equation}
equation \eqref{eq:balance} can be solved with ease. 
In the simplest case of a constant shear modulus ($\mu = \mu_c$), \emph{analytical} solutions can be found via quadrature by similarly projecting the displacement onto harmonics and using Clebsch-Gordon identities. 
For instance, for the dipole component we find
\begin{equation} \label{eq:lagfn}
\begin{aligned}
    \xi^{\rm dip}_{\phi}(r,\theta) =& -\frac{3 \sin\theta\cos\theta}{20 \pi^2 r^3 \mu_c} \times\\
    &
     \Big\{ r^5 \int dr \frac{\Psi_1(t,r) \frac{\partial \Phi_1(t,r)}{\partial r} - \Phi_1(t,r) \frac{\partial \Psi_1(t,r)}{\partial r}}{r^4}  \\
    &+ \int dr r \frac{\partial}{\partial r} \left[ \Psi_{1}(t,r) \Phi_1(t,r)  \right] \Big\},
    \end{aligned}
\end{equation}
which shows effectively that the stress is sourced by poloidal-toroidal cross-terms, while the Maxwell stress is instead sourced by magnetic pressures (which could indeed be much larger). 
Implementing a routine that mimics that described in Sec.~\ref{sec:outbursts} for resetting the crust following an overstraining, we compare three cases here: (i) $|\mathcal{M}_{ij} - \mathcal{M}_{ij}^{\rm eq}|$ for the pure Maxwell stress \cite[as in][]{pons11,perna11,deh20}, (ii) $|\mathcal{M}^{\rm T}_{ij} - \mathcal{M}^{\rm T,eq}_{ij}|$ for the \emph{traceless} component of the Maxwell stress (an adjustment we have made in this paper; see equation \ref{eq: traceless maxwell tensor}), or 
(iii) $|\boldsymbol{\tau}_{ij} - \boldsymbol{\tau}^{\rm eq}_{ij}|$ for the true mechanical strain \eqref{eq:elastic} determined by solving Eq.~\eqref{eq:balance}.

\begin{table*}
\centering
\caption{Relevant details for a variety of (plausibly) isolated LPTs \cite[cf.][]{sdp25b}, namely upper limits on the X-ray luminosity ($L_{\rm X}^{\rm max}$), the observed pulse period ($P$; in descending order) and its (upper-limit) derivative ($\dot{P}^{\rm max}$), together with estimated duty cycles ($D$) and distances ($d$). Additional details on DA J1832 are provided in Tab.~\ref{tab:basics}. Some object names have been shortened. }
\hspace*{-1.6cm}
     \begin{threeparttable}
\begin{tabular}{lccccc} 
\hline
Name &  $L_{\rm X}^{\rm max}$  & $P$ & $\dot{P}^{\rm max}$ & $D$ & $d$ \\
 &  (erg/s) & (s) & (s/s) & ($\%$) & (kpc) \\
\hline
\hline
CHIME J0630+25 \citep{dong24} &  $\sim 10^{29}$ & 421.35542(1) & $ 1.6 \times 10^{-12}$ & $\sim$~0.6 & 0.17(8) \\
GLEAM-X J1627 \citep{hw22} & $2 \times 10^{30}$ & 1091.1690(5) & $1.2 \times 10^{-9}$ & $<$~5 &  1.3(5) \\
GPM J1839--10 \citep{hw23,men25} &  $2 \times 10^{33}$ & 1318.1957(2) & $3.6 \times 10^{-13}$ & $\lesssim  22.7$ & 5.7(2.9) \\
DA J1832$^{\$}$ \citep{li24,wang24} &  $\sim 10^{33}$ & 2656.247(1) & $9.8 \times 10^{-10}$ & $<$~10 & 4.8(0.8) \\
ASKAP J193505.1+214841.0 \citep{caleb24} &  $4 \times 10^{30}$ & 3225.309(2) & $\lesssim 2.7 \times 10^{-10}$ & $\sim$~1.5 & 4.9(5) \\
GCRT J1745--3009 \citep{hyman09} &  $\lesssim 2 \times 10^{34}$ & 4620.72(1.26) & -- & $\sim$~13 & $\sim$~8(?) \\
A J1839 \citep{lee25} &  $10^{33}$ & 23221.7(1) & $1.6 \times 10^{-7}$ & $\lesssim$~3.1 & 4.0(1.2) \\
1E 161348--5055$^{\ast}$ \citep{del06} & $> 10^{33}$ & 24030(108) & $1.6 \times 10^{-9}$ & 0(?) & 3.9(8) \\
\hline
\end{tabular}
\label{tab:isolated LPTs}
        \begin{tablenotes}
\item {\textbf{Notes.} For each source, X-ray limits are quoted assuming the mean distance and `typical' hydrogen column densities, as per the references, at one digit of precision to roughly account for uncertainties.  $\$$ Observed X-ray luminosity during outburst (upper limit $\lesssim 5 \times 10^{31}$~erg in quiescence). $\ast$ Depending on classification, 1E 1613 may not qualify as a LPT as it does not pulse in radio; the source is X-ray volatile, as $L_{\rm X}$ varied by two orders of magnitude between observations spanning 1990 to 2004.}
        \end{tablenotes}
    \end{threeparttable}
\end{table*}

For the sake of demonstration, we choose simple flux profiles that resemble that from \cite{qub25}. 
Explicitly, suppose $\Phi_{1} = B_{p} \left(a_{2} r^2 + a_{4} r^4 \right)$ and $\Psi_{1} = B_{t} \left(b_{2} r^2 + b_{4} r^4 \right)$ where the constants $a_{i}$ are chosen to match to a force-free dipole at the surface and $b_{i}$ are chosen to make the toroidal field vanish there. 
The prefactors $B_{p}$ and $B_{t}$ set the strengths of the poloidal and toroidal components, respectively, which we choose to abide by the phenomenological volume-averaged solution to the induction equation introduced by \cite{Aguilera2008},
\begin{equation} \label{eq:bfield}
    B_{p,t}(t) = B^{0}_{p,t} \frac{e^{-t/\tau_{\Omega}}}{1 + {\tau_{\Omega}}/{\tau_{\rm H}} \left( 1 - e^{-t/\tau_{\Omega}} \right)},
\end{equation}
where $\tau_{\Omega}$ and $\tau_{\rm H}$ are some fiducial Ohmic and Hall times.
We take $\epsilon = 0.85$ (the exact value does not impact the results much) and run a looping script for some choices of $\sigma_{\rm max}$, $\mu_c$, and so on, the numerical values of which are unimportant. 
A time-step is chosen to be small enough ($\Delta t = 10^{-5} \tau_{\Omega}$) that there is at least one step of ``evolution'' through expression \eqref{eq:bfield} between events to ensure accurate counting.

Resulting counts are shown in Figure~\ref{fig:comp} for either a weak toroidal field (left) or one more closely resembling cases studied in the paper with an energy ratio of $E_{\rm tor}/E_{\rm pol} = 0.2$ (right). 
In the case of a weak toroidal field, we see indeed the Maxwell approximation is poor: a factor $\gtrsim 10$ overestimate for the event rate is obtained. 
However, due to the cross-term nature of terms within expression \eqref{eq:lagfn}, as we approach a more equal distribution of energies the discrepancy is reduced, down to order $\sim 10\%$ only after an Ohmic time has elapsed.
Moreover, the traceless criteria is more conservative relative to the full Maxwell case and yields event rates closer to the elastic case. 
If we take larger values of $B^{0}_{p,t}$ more counts are obtained, but the same quantitative trend is observed that cases with $E_{\rm tor}/E_{\rm pol} = \mathcal{O}(1/2)$ give very close agreement between the traceless and elastic ``runs''. 
Again we emphasise this is a simple demonstration, but suggests that the results obtained here are broadly consistent between the three plausible failure criteria.

For transparency, we close this section by noting that there are some important caveats worth highlighting with respect to our models. 
Probably the most critical aspect is that there is no explicit inclusion of elasticity or plasticity: 
failures are handled through an instantaneous, von Mises mechanism where the crust heals prior to the next time-step without undergoing any lateral motions. 
In particular, while we have used the von Mises criterion to determine failure sites, molecular dynamics simulations instead indicate that a more physical mechanism is the Zhurkov model \citep{chugunov10}, where failures occur when thermodynamic fluctuations exceed a threshold energy over finite time intervals. 
This leads to gradual deformations rather than immediate failures and is more appropriate for modelling elastic/plastic transitions, though is difficult to handle numerically owing to the timescales of the problem. 
Having the crust reset instantly is also not really consistent with the radio-activation picture, since the zone should remain plastic for multiple time steps (and hence not accumulate stress) to account for a non-negligible duty cycle \citep{coop24}. 
Straining motions should also contribute a Faraday-like $\boldsymbol{v}_{\rm ion} \times \boldsymbol{B}$ term within equation \eqref{eq: induction equation}  \citep{lg19,brans20}, though we do not explicitly account for this feature in this first magnetothermal study of isolated LPTs. 

\section{Relevant properties of isolated LPTs}  \label{sec:somelptproperties}

Details on data used in the main text to support the `late-bloomer' position are provided here. 
Table~\ref{tab:isolated LPTs} gives X-ray upper limits ($L^{\rm max}_{\rm X}$), spin period, period derivative, observed duty cycles, and distances, respectively, for eight LPTs that appear isolated. 

An object deserving of individual discussion is CHIME J0630. 
The low dispersion measure (DM; $\approx 22(1) \text{ pc cm}^{-3}$) of this source indicates it may be as close as \citep{dong24} $\gtrsim 70$~pc at $95\%$ confidence. 
This effectively rules out a fallback disk due to the absence of optical, NIR, and UV counterparts. 
Interestingly, trace contents of the radioactive iron nuclide $^{60}$Fe found in deep-sea archives of Earth's oceans reveal interstellar influxes some time between $1.5$ and $3.2$ Myr ago from nearby supernovae \citep{wall16}. 
Such an age fits naturally within the context of models presented here. On the other hand, pulse timing from CHIME J0630 indicates a period of only $P \sim 7$~min, suggestive of inefficient failure-to-spindown conversions; this can be accommodated within our model for $\tilde{\zeta} \lesssim 8 \times 10^{-4}$ (see Fig.~\ref{fig:ct_sd}) or a weaker field. 
Owing to its proximity, this object should be prioritised for X-ray monitoring where even shallow bursts should be visible, especially given the likelihood of a timing glitch in May 2021.
At the lower limit of the DM-inferred distance, the upper-limit to the persistent X-ray luminosity \citep{dong24} reads $\approx 4 \times 10^{28} \text{ erg s}^{-1}$---even harder to explain with CC models.

GPM J1839 has tight timing constraints \citep{men25} ($\dot{P} \lesssim 3.6 \times 10^{-13} \text{ ss}^{-1}$), spanning the last $\sim 3$~decades, set while the source was intermittently radio-loud. 
Owing to the mild limits on this source's X-ray luminosity \citep{men25}, shallow failures triggering radio activity \citep{coop24} remain plausible provided the present-day field is comparatively weak ($\lesssim 10^{14}$~G) and that the bulk of the enhanced spindown occurred prior to the present epoch when the magnetoelastic energy reservoir was plentiful. 

\subsection{DA J1832 activity timeline and alternative scenarios} \label{sec:daj1832}

Aside from LPTs generally, we have discussed the case of DA J1832 at length since this object has displayed both X- and radio-band activity simultaneously. 
We provide here a brief summary of the observational timeline and how we have interpreted it theoretically, providing also some caveats and alternative interpretations. 

The sky surrounding DA J1832 has been regularly monitored over the last $\sim$~decade because of the coincident position of the supernova remnant G22.7--0.2. 
No pulsations had ever been recorded, despite $\sim$6000 hours of archival data having been accumulated by the Very Large Array since 2017. 
Radio pulses at a consistent period of $P = 2656.247$~s (derivative upper limit $\dot{P} < 10^{-9} \text{ ss}^{-1}$) were subsequently detected by ASKAP with brightness $\approx 1.9$~Jy on 8 December 2023. 
The source was highly variable (spanning $\sim$~20~mJy to 18~Jy) through to February 2024 at frequencies ranging from $\approx$~0.3 to 3~GHz, at which time a serendipitous X-ray search by Chandra measured an unabsorbed flux of $\sim 5 \times 10^{-13} \text{ erg s}^{-1} \text{ cm}^{-2}$ in the 1--10 keV band, showing no signs of decay during the $20$~ks over which photons were intercepted. 
Since $\text{DM} \approx 458 \pm 14 \text{ pc cm}^{-3}$, a distance of $d= 4.5^{+1.2}_{-0.5}$~kpc is expected based on electron map models and foreground gas measurements, implying a peak luminosity of $L_{\rm X} \sim 10^{33} \text{ erg s}^{-1}$ for a hydrogen column density of $N_{\rm H} = 1.8 \times 10^{22} \text{ cm}^{-2}$. 
Spectral fits support a power-law component together with a $\gtrsim$~keV hotspot of a (notably small) radius $\lesssim 100$~m. 
Followup searches in August 2024 by Chandra and the Einstein Probe placed upper limits of $L_{\rm X} \lesssim 7 \times 10^{31} \text{ erg s}^{-1}$, though the source remained radio-loud albeit dim through September 2024 at $\sim 60$~mJy.

\begin{figure}
\centering
  \includegraphics[width=0.497\textwidth]{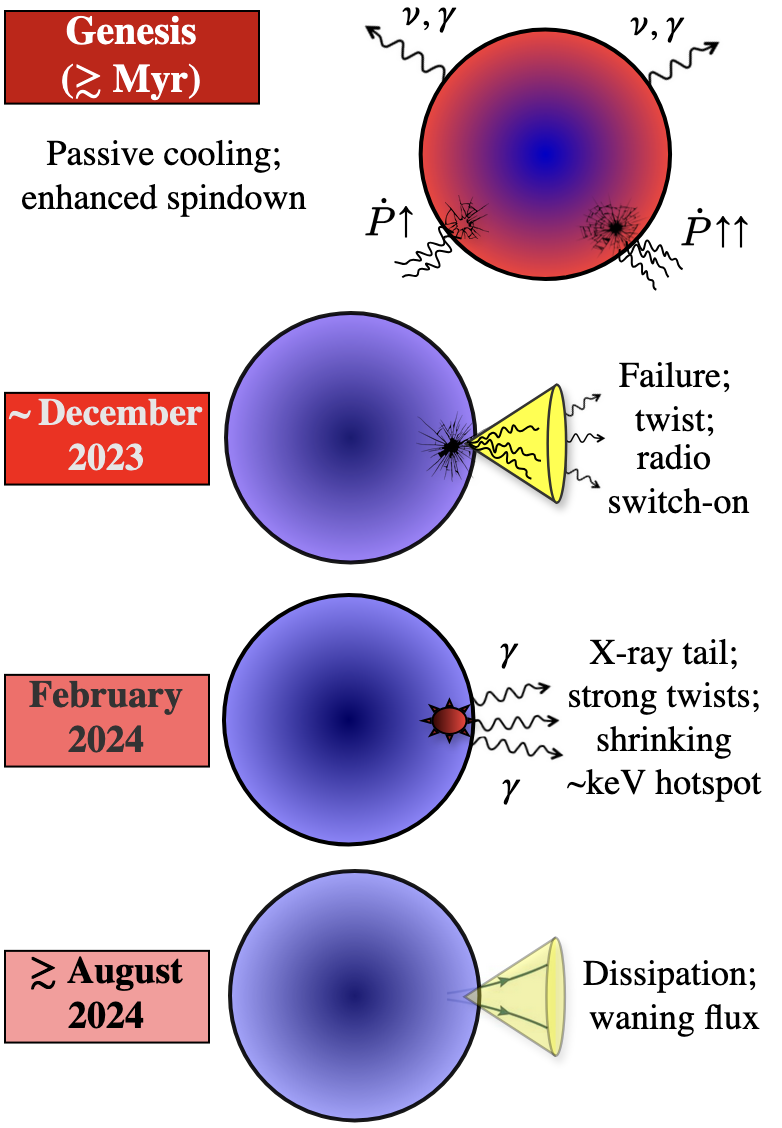}
  \caption{Timeline of DA J1832. (i) A silent era of neutrino and photon cooling whereby the electrical conductivity gradually increases, eventually instigating late-time activity and rapid spindown; (ii) crustal failure event in December 2023 and onset of radio pulsing; (iii) X-ray dissipation of the spot, the tail of which was observed in February 2024; (iv) twist decay and eventual shutoff.}
  \label{fig:timeline}
\end{figure}

Note in particular that X-ray data were unavailable during the onset of radio pulsing in December 2023. 
Although one could take the observations at face value (see below), since the absence of evidence is not evidence of absence we have instead hypothesised the following: a (sub)-crustal event took place around the start of December 2023, which was directly responsible for radio activation by injecting magnetospheric twist or initiating a plastic flow. 
Thermoplastic heating or particle backflow \citep{bell14} generated a localised hotspot surrounding the failure site as the twist ramped before decaying in the $\lesssim$~months following February 2024. 
A schematic illustrating this scenario is depicted in Fig.~\ref{fig:timeline}, while a summary of relevant source parameters are listed in Tab.~\ref{tab:basics}. 

\begin{table}
	\centering
	\caption{Summary of relevant observed properties of DA J1832. Here we have introduced the radio flux density $S_{\nu}$, the considerable variation of which may be attributable to a waning twist or magnetospheric pollution from fireball plasma.
}
\hspace{2pt}
\begin{tabular}{cc}
	    \hline
	    Observed/derived properties & Value \\
	    	    \hline
	    	    	    \hline
	    Spin period ($P$) & $2656.247 \pm 0.001$~s \\
	    Period derivative ($\dot{P}$) & $< 10^{-9} \text{ ss}^{-1}$ ($95\%$) \\
	    $L^{\rm quiescent}_{\rm X}$ (ca. 2011) & $< 5.3 \times 10^{31}$~erg/s ($3 \sigma$) \\
	    $L^{\rm quiescent}_{\rm X}$ (Aug. 2024) & $< 6.5 \times 10^{31}$~erg/s ($3 \sigma$) \\
	    $L^{\rm active}_{\rm X}$ (Feb. 2024) & $\sim 10^{33}$~erg/s \\
	    $S^{\rm quiescent}_{\nu}$ (2017--2023) & $< 10$~mJy \\
	    $S_{\nu}$ (Dec. 2023) & $\sim 1$~Jy \\
	    $S_{\nu}$ (Feb. + Mar. 2024) & $\sim 10$~Jy \\
	    $S_{\nu}$ ($\sim$~Sep. 2024) & $\sim 60$~mJy \\
     \hline
	    \hline
	\end{tabular}
	\label{tab:basics}
\end{table}

Interestingly, \cite{li24} suggested that there is an association between SNR G22.7--0.2 and DA J1832. 
However, this seems improbable because 
(i) Sedov-Taylor theory of expansion predicts an age of $\sim 25$~kyr, while
(ii) for such an age, the location of DA J1832 indicates it must have been travelling some $\sim 400$~km/s since birth (i.e., natal kick) but 
(iii) VLBI measurements (possible because of the multi-instrument radio detections) place an upper limit of $\sim 190$~km/s. 
Even ignoring troubles with age therefore, an association seems unlikely as concluded by \cite{wang24}.

Aside from this, the minimal timeline model considered in Fig.~\ref{fig:timeline} does not take the observations at face value: the source started pulsing in radio in November 2023, exhibited heightened X-ray activity in February 2024, and then shut-off in X-rays but continued pulsing for a while in radio. 
Such a timeline is unusual for magnetars. 
Almost always there is a burst \emph{before} radio activation, though in some rare cases (such as 1E 1547.0--5408 in 2022 and PSR J1119--6127 in 2016) radio activity is shut off some \emph{weeks} prior to a (relatively weak) outburst \citep{low23}. 
In these latter objects, however, it is plausible that early-time crustal activity went undetected due to the high baseline X-ray flux from the sources and so perhaps they are, in fact, not unusual.

Envisioning a situation where the source turns on in radio and \emph{then} becomes X-ray visible appears to require a magnetospheric origin where the field geometry becomes conducive to radio pulsing through an instability or other means. 
One may imagine that there is a backreaction realised at the surface such that local and some global heating occurs which then dissipates faster than the magnetospheric twist. 
This could, in principle, source radio activation \emph{followed} by X-rays, with persistent but waning radio as observed. 
This appears difficult from an energetics perspective though. A spot produced due to backflow has an area of
\begin{equation}
A = \pi R^2 u,
\end{equation}
where $u$ effectively determines the size of the current-carrying bundle where particles are created due to enhanced twist. 
For DA J1832 with a spot of size $\sim 100$~m we have $u \sim 10^{-4}$. The anticipated luminosity of the spot due to kinetic heating reads \citep{belo09}
\begin{equation}
L \approx 1.3 \times 10^{28} \times B_{14} R_{6} \psi V_{9} u_{-4}^2 \text{ erg s}^{-1},
\end{equation}
where $V_{9}$ is a threshold voltage (in units of $10^{9}$~V), defined as the value of the electrostatic potential $\Phi_{e}$ whereupon copious particle supply is available to carry large currents, and $\psi$ denotes the toroidal twist. 
The luminosity decay timescale in this picture reads \citep{belo09}
\begin{equation}
t_{\rm ev} \sim -\frac{L}{dL/dt} \approx 15 \times V_{9}^{-1} B_{14} R_{6}^2 \psi u \text{ yr}.
\end{equation}
Matching to observations with $t \sim 6$~months to eliminate $V_{9}$ gives a maximum luminosity of
\begin{equation} \label{eq:lumeq}
L_{\rm max} \sim 10^{26} \times B_{14}^2 \text{ erg s}^{-1}.
\end{equation}
Even for ultra-strong fields, the luminosity \eqref{eq:lumeq} is $\sim 7$ orders too low to match observations of DA J1832. 
We may thus exclude backflow heating from particle bombardment. 

\section{Burst-to-spindown efficiencies} \label{sec:sdcontext}

To provide context for the value of $\bar{\zeta} \sim 10^{-3}$ stipulated in Sec.~\ref{sec:sd}, we consider some relevant observations and theoretical elements here. 
We first point out that, regrettably, the failure energy -- that which is accessible from a simulation standpoint -- is not directly observable and must instead be inferred from X-ray data. 
In general, we may decompose
\begin{equation}
\bar{\zeta} = \zeta_{1} \zeta_{2},
\end{equation}
where $\zeta_{1}$ represents an intermediate conversion factor between (magneto-)elastic and outburst energies, with $\zeta_{2}$ then being the fraction of outburst energy that contributes to spindown.
Note that our $\bar{\zeta}$ is phenomenologically similar to that used in other studies of enhanced spindown \cite[see, e.g., figure 1 in][]{ben20}.

The global simulations by \cite{brans20} of pulsar quakes where the crust and core are coupled -- as appropriate to CT models -- suggest that $\sim 1\%$ of the quake energy escapes as observable radiation (i.e., $\zeta_{1} \sim 10^{-2}$). 
This radiation escapes as waves which may not directly bolster the dipole moment, but gives us a handle on how to relate numerically-determined quake energy to observable radiation. 
For magnetars, $\zeta_{1}$ may be higher as the ambient magnetic field mitigates thermodynamic losses:
\cite{blaes89} calculated the crust-to-magnetosphere transmission coefficient, finding it scales with the field strength at high frequencies, especially in cases where the affected region liquefies. It could reach $\sim \mathcal{O}(10\%)$ for $B \gtrsim 10^{14}$~G.

Consider now some observations of bursts of relatively low luminosity from magnetars. 
For instance, 1E 1547.0--5408 released an X-ray burst in 2022; integrating the event light curve, the total burst energy can be estimated \citep{low23} as $1 \lesssim E_{\rm X}/\left( 10^{41} \text{ erg}\right) \lesssim 6$ accounting for uncertainties on column density and distance. 
Relative to the `baseline' spindown rate of $\dot{\nu} = -3.63(3) \times 10^{-12} \text{ s}^{-2}$, an \emph{average} net increase of $|\delta \dot{\nu}| \approx 2.54 \times 10^{-12} \text{ s}^{-2}$ was observed  over $\approx 147$ days following the event \citep{low23}.
This roughly translates into an excess rotational kinetic energy loss of $\gtrsim 3 \times 10^{41}$~erg. 
For this, possibly extreme, event we could thus anticipate $\zeta_{2} \sim \mathcal{O}(1)$ and hence $\bar{\zeta} \sim \mathcal{O}(10^{-2})$. 

Perhaps more appropriate for LPTs, GPM J1839--10 exhibits linear-to-circular polarisation \citep{men25} conversion in a way that is reminiscent of XTE J1810--197, suggesting their magnetospheric environments may be alike. 
The latter magnetar is known to enter into epochs of enhanced spindown post outburst. 
Following its 2003 event(s), $\dot{\nu}$ increased by a factor $\approx 6$ that stabilised over the next $\sim 3$~years \citep{cam16}. 
Spindown remained relatively steady until 2018 when the source burst again, with an average increase of $\delta \dot{\nu} / \dot{\nu} \approx 2.5$. 
Pulsed fraction differences point towards different emission geometries between events \citep{borg21}, which could be responsible for the discrepancy in $\delta \dot{\nu}$ increases across events. 
The burst energy can be further estimated as $\sim (4 \pm 2) \times 10^{42}$~erg while the source lost $\sim 10^{41}$~erg of rotational kinetic energy, suggesting $\zeta_{2} \sim \mathcal{O}(0.1)$ and hence $\bar{\zeta} \sim \mathcal{O}(10^{-3})$.


\end{document}